\documentclass[12pt,twoside,a4paper,fleqn]{article}
\usepackage[left=25mm,top=20mm,right=14mm,bottom=25mm]{geometry}

\usepackage{epsfig}
\usepackage{graphicx}
\usepackage{amssymb}
\usepackage{mathrsfs}
\usepackage{dcolumn}

\usepackage{multirow}% http://ctan.org/pkg/multirow

\newcommand{\pr}{\partial}

\newcommand{\rta}{\rightarrow}

\newcommand{\om}{\omega}

\newcommand{\beq}{\begin{equation}}
\newcommand{\eeq}{\end{equation}}

\newcommand{\ball}{\begin{align}}
\newcommand{\eall}{\end{align}}

\newcommand{\beqar}{\begin{eqnarray}}
\newcommand{\eeqar}{\end{eqnarray}}

\newcommand{\ben}{\begin{enumerate}}
\newcommand{\een}{\end{enumerate}}
\makeatletter
\newcommand*{\rom}[1]{\expandafter\@slowromancap\romannumeral #1@}
\makeatother

\begin{document}
\date{}
\title{A Brief history of mangnetism}
\author{ Navinder Singh$^*$ and Arun M. Jayannavar$^{**}$ \\ $^*$Physical Research Laboratory, Ahmedabad,\\ India, Pin: 380009.\\$^{**}$ IOP, Bhubaneswar, India.
%\footnote{~Email:~sharmakomal611@gmail.com, komal.phyhpu@gmail.com}
	\footnote{ ~Email:~navinder@prl.res.in}
\footnote{~Email:~jayan@iopb.res.in }
}
\maketitle
\begin{abstract}
In this article an overview of the historical development  of the key ideas in the field of magnetism is presented. The presentation is semi-technical in nature.Starting by noting down important contribution  of Greeks, William Gilbert,  Coulomb, Poisson, Oersted, Ampere, Faraday, Maxwell, and Pierre Curie, we review  early 20th century investigations by Paul Langevin and Pierre Weiss. The Langevin theory of paramagnetism and the Weiss theory of ferromagnetism were partly successful and real understanding of magnetism came with the advent of quantum mechanics.  Van Vleck was the pioneer in applying quantum mechanics to the problem of magnetism and we discuss his main contributions:
(1) his detailed quantum statistical mechanical study of magnetism of real gases; (2) his pointing out the importance of the crystal fields or ligand fields in the magnetic behavior of iron group salts (the ligand field theory); and (3) his many contributions to the elucidation of exchange interactions in d electron metals. Next, the pioneering contributions (but lesser known) of Dorfman are discussed. Then, in chronological order, the key contributions of Pauli, Heisenberg, and Landau are presented. Finally, we discuss a modern topic of quantum spin liquids.
\end{abstract}
%%%%%%%%%%

\section{Prologue}

In this presentation, the development of {\it the conceptual structure of the field is highlighted}, and the historical context is somewhat limited in scope. In that way, the presentation is not historically very rigorous.  However, it may be useful for gaining a "bird's-eye view" of the vast field of magnetism.

\section{Ancient Greeks to medieval times}
%—————————————————————————————————————————

The "magical" properties of magnets have fascinated mankind from antiquity. A naturally occurring magnet,  lodestone\footnote{Lodestone is a naturally magnetized lump of the mineral magnetite (an oxide of iron). Earth's magnetic field is too weak to magnetize it. It is probably lightning bolts that magnetize it, as lodestone is found near the surface of earth, not deep underneath.}, was known to ancient Greeks. The famous Greek philosopher Thales of Miletus documented loadstone's magnetic properties in the 6th century BC\cite{shoen}. In the same century the ancient Indian physician Sushruta was aware of the magnetic properties of loadstone, and used it to remove metal splinters from bodies of injured soldiers\footnote{His work "Sushruta-Samhitha" has been described as the most important treatise on medicine and surgery in ancient times and is the founding work of Ayurveda\cite{anne}}. However, according to reference\cite{yang}, ancient Chinese writings that date back to 4000 BC mention magnetite and use of magnetic compass in navigation. Apart from the question of the discovery of loadstone, it is clear that it was known in the BC era, and was used for the benefit of mankind whether in surgery or in navigation.\footnote{It is interesting to note that the phenomenon of frictional electricity was also discovered in that era and these two subjects developed largely independently from each other until these were "wedded" together in a wider subject of electromagnetism in the 19th century, by the investigations of Faraday, Maxwell, and others\cite{r1,r2}.}    
 
However, before William Gilbert of England in 17th century,  study of magnetism was meta-physical. Along with loadstone's good use many superstitions were attached to it (due to its "magical" magnetic properties). It was believed that loadstone feeds on iron, an iron piece kept near loadstone would loose its weight while loadstone would gain. But falsey  of such meta-physical claims was clear when the experimental method of science started in 16th century, and experiments to test this "weight gain-weight loss'' effect gave negative results. Another popular superstition attached to loadstone was that it can act as a pain killer, and it was used as an amulet to be tied around affected body parts. Today we know that these beliefs have no scientific basis and magnetism has no known direct effect on biological or psychological processes.

%------------------------------------------------
\section{Enter Gilbert}   
%%———————————————————————————

Scientific study of magnetism started in the 16th century with the investigations of William Gilbert (1544 - 1603) of England. He studied medicine and started his practice as a physician in 1573. He lived in good times of the Elizabethan renaissance in music, art, and in Natural philosophy, and was private physician of the queen Elizabeth I from 1601 to 1603. Along with medicine did his pioneering investigations in natural philosophy (as  subject of science was called in those times). \footnote{He supported the philosophy of Nicolus Copernicus (1473 - 1543) that earth is not the centre of the universe, and rejected the old Aristotelian philosophy.} The most important work of his life is "De Magnete'' in which he summarized all his investigations of magnetic and static electricity phenomena. He clearly differentiated between static electricity and magnetism. He also studied the effect of temperature on magnetic properties of iron and discovered that when iron is red hot it ceases to be attracted by a magnet\cite{r1}.  {\it His most important discovery is the realization that earth itself is a large magnet and  North-South pointing property of compass needle can be understood from the attraction of unlike poles of magnets thereby bringing the mysterious pointing properties of compass needle to correct scientific explanation}.

%------------------------------------------------

\section{Enter Coulomb}
%%%—————————————————————

Then 17th century was sort of long lull regarding the subject of magnetism, not much was done in this field. It is interesting to note that great Issac Newton (1642 - 1726) did not contribute much to magnetism, although he developed all the other subjects significantly, and created many new, as is well known. It could be possible that he thought about  cause of magnetism and could not reach on to experimentally testable concrete conclusions. His account of magnetism in his greatest work  "Principia'' is very ambiguous\cite{principia}.

18th century saw a major development. Charles-Augustin de Coulomb (1736 - 1806) experimentally found the inverse square force law (now famously known as the Coulomb law) using his torsion balance set-up. He started his career as an engineer in French army in 1761, and continued his service in next two decades at various places in France, and one long assignment outside France in West Indies. In the decade 1780 to 1790 he devoted himself to the study of electric and magnetic phenomena, and experimentally deduced the force laws between charges and magnetic poles.\footnote{It turns out that John Mitchell (1724-1793) who was contemporary of Coulomb discovered torsion balance, and found inverse square force law between magnetic poles prior to Coulomb\cite{r1}. However, Coulomb performed much more systematic and detailed studies, and wrote about it extensively.} It is interesting to note that in those days the theoretical understanding of electric and magnetic phenomena was a sort of meta-physical. It was believed that there is an invisible fluid that carriers magnetic and electric effects from one place to another. This was known as the "effluvia theory".\footnote{This state of affairs is much like that in the Caloric theory of heat in which it was believed that heat is a kind of fluid (the Caloric) which flows from hot bodies to cold bodies. Later on, it was experimentally proved wrong by Count Rumford who observed that large amount of heat is created when iron rods are drilled to make gun barrels, this violated the conservation of Caloric, and mechanical theory of heat was advanced.} The effluvia theory was discarded much later, in the 19th century, when field concept was originated by Faraday. It is also interesting to note that Coulomb believed in the effluvia theory but performed accurate experiments. And these are his accurate experimental findings that stood the test of the time and immortalized his name!

%------------------------------------------------
\section{Enter Poisson}
%%————————————————————
Next major development in the field of magnetism was brought by Simeon Denis Poisson (1781-1840) and Carl Friedrich Gauss (1777 - 1855). These men formulated Coulomb's experimental findings into an elegant mathematical theory.  Poisson obtained mathematical expressions for force law for an arbitrarily shaped magnetic material in terms of surface and volume integrals. Calculus was extensively used to analyze these problems. The concept of "potential'' was invented for quantitative analysis of problems related to electric and magnetic phenomena.\footnote{Poisson was a leading opponent of Huygen's wave theory of light even though in 1800 Thomas Young convincingly demonstrated wave nature of light using his two slit experiment. Poisson strongly believed in Newton's corpuscular theory and also opposed Jean Fresnel's wave - theoretical explanation of diffraction.}

\section{Enter Oersted}
%%—————————————————————
After the invention of the first electric battery in 1800 (the Voltaic pile) a major discovery happened in the field of electricity and magnetism which would bridge the gap between the two. In 1820, Danish physicist Christian Oersted (1777 - 1851) discovered that electric current produces magnetic effects. In the famous experiment with his lab assistant Hansteen he observed that a current carrying wire has the capacity to deflect a nearby placed magnetic needle. However, this discovery was not totally unexpected as Oersted believed that there might be a connection between electricity and magnetism. His belief of this intimate connection was a result of influence on him of Immanuel Kant's philosophy which he studied for his dissertation in 1799. Kantian philosophy asserts deep connections in natural phenomena and unity of Nature.

\section{Enter Ampere}   
%——————-———————————————
The French scientist Andre - Marie Ampere (1775 -1836) took it further and developed the idea that "magnetic force'' around a wire has circular character and it leads to a law which is now called the Ampere law (written down in a mathematical form by J C Maxwell). He also performed further experiments on magnetic effects of currents like measuring forces between current carrying wires.  Other French investigators of his time (in this field) were Biot, Savart, and Arago.  They also contributed to the development of the field\cite{mattis}. {\it Perhaps his most important contribution with respect to magnetism of matter is his explanation that Ferromagnetism is a result of internal currents in these materials.} This is much close to present day understanding. He was aware of Joule heating effects and ascribed the internal currents to "molecules" of iron as giving magnetic effects. Notice that notion of molecules and atoms at that time was not well established.  Thus most important contributions of Ampere are his {\it Ampere's law} and his explanation of ferromagnetism from microscopic circulating currents. 

\section{Enter Faraday}
%———————————————————
One  of the most important name in the development of the field of magnetism is Michael Faraday (1791 -- 1867)--the self-trained experimental genius of 19th century. Born in a family which was not well to do financially, Faraday had to struggle and had only very basic school education. He studied mostly by himself. At a young age of 14 he became an apprentice to a local book binder and seller. This gave an opportunity to read, and Faraday read most of the books that came to him for binding. The second lucky break came to him when the eminent English chemist Humphry Davy of the Royal institution supported him and offered a job of lab assistant in his lab. This started Faraday's scientific career. In 1831 Faraday discovered the law of electromagnetic induction -- arguably the most important discovery of the 19th century. Faraday was motivated by his studies, in collaboration with Sir Charles Wheatstone on vibrational phenomena in iron plates, in which {\it acoustic induction} caused one plate to vibrate when a nearby plate is vibrated by an external means. It seems that this acoustic induction is behind Faraday's most famous experiment of the "electromagnetic'' induction.  He wound a coil of insulated copper wire around a thick iron ring on one side, and an other coil on the opposite side. He connected one of the coils to a galvanometer, and the other to a voltaic pile through a switch. When the switch was closed or opened he observed sudden deflection the galvanometer. This marks the discovery of the phenomenon of electromagnetic induction. In the fall of the same year 1831 Faraday also noticed that a current can be induced in a coil by moving a permanent magnet close to it. We now know how important these discoveries are in the development of the filed of electricity  and magnetism, and how these discoveries lead to the technological revolution.

Thus Faraday took the studies of Oersted and Ampere forward and developed the subject of electromagnetism. Most importantly in the theoretical understanding of the phenomena Faraday advanced the field of electromagnetism from metaphysical conceptions of {\it{magnetic fluids}} (effluvia theory) to the modern concept of {\it{magnetic and electric fields}}.  The field concept and the concept of lines of force are his one of the biggest contributions.  With the concepts of "magnetic field'' and "lines of forces'' he could explain these phenomena, and latter on Maxwell developed the mathematical theory of electromagnetism based on Faraday's conceptions. Along with his pioneering experiments in electricity and magnetism Faraday also discovered the field of electrochemistry. As this brief historical introduction is aimed at the historical development of magnetism, we skip these interesting discoveries and interested reader can read about these in a dedicated biography of Faraday, for example in\cite{faraday}.
      
Before him only strongly magnetic materials were known like iron and cobalt which shows ferromagnetism (a strong form of magnetism). Faraday also discovered "weak magnetism'', i.e., dia- and para-magnetism in substances like oxygen and bismuth. He observed that paramagnetic substances are attracted towards stronger magnetic fields while diamagnetic substances are repelled.

\section{Enter Maxwell}
%——————————————————————
A consolidation of Faraday's experimental findings was achieved by James Clerk Maxwell (1831-1879) who expressed those findings by an elegant mathematical language. The result was the famous Maxwell equations. With his profound intuition he was able to go beyond Faraday and introduced his "displacement term''.  His theoretical synthesis was primarily based on Maxwells investigations. The following lines by Maxwell shows how closely he followed Faraday:

".........resolved to read to no mathematics on the subject till I had first read through Faraday's Experimental researchers in Electricity. I was aware that there was supposed to be a difference between Faraday's way of conceiving phenomena and that of the mathematicians.........As I proceeded with the study of Faraday, I perceived that his method.......capable of being expressed in ordinary mathematical forms.  For instance, Faraday, in his mind's eye, saw centers of force traversing all space where the mathematicians saw centers of force attracting at a distance: Faraday saw a medium where they saw nothing but distance.\cite{max}"

In a nutshell, Maxwell tied the final remaining lose threads of the unification of electricity and magnetism and the field of electromagnetism was born. This further enabled Albert Einstein to point out fundamental problems in classical mechanics of Issac Newton and lead to the advent of the special theory of relativity. The technological revolution brought about by Maxwell's work is well known. Without going too far away from our main topic of the history of magnetism we stress here that the findings of Coulomb, Oersted, Ampere, Poisson, Faraday, and others were expressed in a beautiful mathematical language by Maxwell which lead to further developments in magnetism and other fields.

\section{Enter Pierre Curie}
%———————————————————————————

Pierre Curie (1859 -1906) advanced further the experimental findings of Michael Faraday on weakly magnetic substances i.e., paramagnetic and diamagnetic materials. He undertook the study of these materials for his doctoral thesis with the aim to investigate whether there are transitions between various kinds of magnetism in a given material. He performed a thorough study of magnetic properties of some twenty substances\cite{curie}. These painstaking investigations leads to three important discoveries: (1) paramagnetism in various salts is temperature dependent and magnetic susceptibility (ratio of induced magnetism to applied magnetic field) is inversely proportional to temperature (known as the Curie law ($\chi \propto \frac{1}{T}$)); (2) Ferromagnetism is also a function of temperature and completely vanishes when  temperature is raised above a critical temperature called the Curie temperature ($T_c$); and (3) diamagnetism is approximately temperature independent. To measure the magnetic coefficients he designed and perfected a very sensitive torsional balance that could measure up to 0.01 mg!

Pierre Curie, along with his wife madam Marie Currie, discovered radium and polonium and did the pioneering investigations in the field of radioactivity. In 1903 they, along with Henri Becquerel, were awarded with Nobel prize for their investigations in radioactivity.  But without any doubt Curie's experimental investigations of weakly and strongly magnetic substances are major contributions to magnetism.  However,  microscopic understanding of various forms of magnetism remained unclear.

\section{Enter Langevin}
%———————————————————————

The experimental studies of Faraday and Curie on weakly magnetic substance remained completely unexplained up to the end of 19th century. The first "microscopic'' understanding of the behavior of diamagnetic and paramagnetic substances came with the investigations of Paul Langevin (1872 - 1946). Langevin, using then newly-discovered statistical mechanics of Boltzmann and Gibbs came up with a mathematical theory which showed that paramagnetic susceptibility is inversely proportional to temperature, whereas the diamagnetic one shows temperature independent behavior. At the core of the Langevin theory was his phenomenological introduction of the tiny magnetic moment of atoms. 

At the time when Langevin advanced his theory (in 1905) atomic structure was unknown (Rutherford's alpha scattering experiment (1911) and Bohr's model of the atoms came many years later (1913)). However, Langevin's ad-hoc introduction of tiny atomic magnetic moments was more or less correct. We now know that magnetic moments in atoms originate either from the orbital motion of unpaired electrons, or from the spin of electrons, or some vectorial combination of these two. For paired electrons magnetic effects of orbital motions nullify each other, and two paired electrons in an orbital give zero net spin by Puali's exclusion principle, thus no magnetic effects. All noble gases in the right most column of the periodic table are examples of this nullifying effect and they are thus not paramagnetic. They show only diamagnetism.

With this ad-hoc assignment of a tiny magnetic moment to each atom of a paramagnetic substance and by including the thermal agitation of magnetic moments Langevin was able to obtain the correct temperature dependence of magnetization of a sample. At very low temperatures, when thermal agitation is feeble, the tiny atomic magnetic moments tend to align along the direction of an applied magnetic field thus showing paramagnetic effect. At higher temperatures due to thermal agitation lesser atomic magnetic moments align along the direction of external magnetic field leading to a weaker paramagnetic effect. It turns out that at the average magnetic moment per unit volume of the whole sample (called magnetization) is given by

\[M = \frac{n \mu^2}{3 k_B T} H,\]  

Where $\mu$ is the tiny magnetic moment of each atom, $n$ is the number density of atoms, $k_B$ is the Boltzmann constant, $T$ is the absolute temperature, and $H$ is the applied magnetic field\cite{Feynman vol 2}. Thus magnetic susceptibility ($\chi = \frac{M}{H}$) is inversely proportional to temperature in accordance with observations.  This is one of the pioneering achievement of Langevin. On similar lines Langevin was able to show that diamagnetic effects occur in substances in which there are no atomic magnetic moments ($\mu = 0$). Although in this case atoms appear magnetically neutral, but there could be induced magnetic effects due to the internal motions of electrons within each atom. Using Faraday's induction law, he was able to show an opposing magnetic effect i.e., the induced magnetization in opposite direction to the applied magnetic field, and from statistical mechanics obtained it's temperature independence in accordance with experiments. Thus by 1905, a partial understanding of weak magnetism (para- and dia- magnetism) was achieved. The origin of the tiny magnetic moments was a mystery at that the time, and another important problem of ferromagnetism remained unresolved.

\section{Enter Pierre Weiss}
%—————————————————————————

Pierre Weiss (1865-1940) generalized the Langevin theory by introducing a new and very important concept of "mean molecular field'' and with his "generalized theory'' he could account for ferromagnetism (however, only partly).

The Weiss theory goes like this: Suppose that a ferromagnetic material (like iron) is placed in an external magnetic field $H$. The field acting on each atom of iron, according to Weiss, is not given by the external field alone but an additional field (a molecular field as Weiss called it) also acts on each atom. This molecular Feld is proportional to the magnetization of the sample. Thus the field acting on each atoms is $H+\lambda M$ instead of $H$. Here $M$ is the magnetization of the sample and $\lambda$ is a material dependent constant. This is the main point of the Weiss theory. The temperature dependence of magnetization is much more stronger in ferromagnetic case and can be qualitatively explained in the following way. When temperature is lowered more and more spins align along $H$ and thereby enhancing $M$. Increased $M$ reinforces the alignment process through the field $H + \lambda M$ until at sufficiently low temperatures saturation sets in, in which all spins point along $H$. Similarly, increasing temperature leads to more and more de-stabilization of the spin alignment thereby reducing $M$. The field $H +\lambda M$ acting on each spin further weakens and at sufficiently high temperatures and in vanishingly small external field $H$ it leads to zero magnetization. So temperature dependence of magnetization is much more stronger in the case of ferromagnetism, as compared to para- and dia- magnetism. It turns out that at sufficiently low temperatures it is energetically favorable for spins to align in one direction, thus there can be a molecular field alone. The Weiss theory can be mathematically expressed if we change $H$ to $H +\lambda M$ in the above Langevin equation:

\[M =  \frac{n \mu^2}{3 k_B T}(H +\lambda M).\]      

Thus leading to

\[M = \frac{n \mu^2/3 k_B}{T- \lambda (n \mu^2/3 k_B)}H.\]

This is known as the Curie-Weiss law. If $T$ is much greater than $T_c = \lambda\frac{n \mu^2}{3 k_B}$ it gives a stronger paramagnetism. Thus Weiss theory qualitatively explained Curie's experimental results for Ferromagnetism. However, it has a serious problem. When experimentally known quantities ($T_c,~~n$ etc) are inserted in the above expression, it turns out that $\lambda \sim 10^{4}$! Within the semiclassical theories it is not possible to explain such a large value of $\lambda$. Thus Weiss theory could only partly shed light on the origin of ferromagnetism. Such large value of $\lambda$ could only be explained with the advent of quantum mechanics, and in 1926 Werner Heisenberg was able to explain such large values of $\lambda$ using the concept of quantum mechanical exchange (refer to article II in this series and references therein).

Also Weiss theory is a classic example of a mean-field theory. When $T\rta T_c$, susceptibility diverges, and the actual scaling law of susceptibility with temperature ($T-T_c$) can only be obtained by going beyond the mean-field approximation. In summary, the above Weiss expression could explain Curie's experimental observations in ferromagnetic substances at $T>T_c$, and from the Weiss expression, negative magnetization for $T<T_c$ means that molecular field is strong enough to spontaneously magnetize a ferromagnetic sample without an external applied field. %Thus all ferromagnetic substances should show spontaneous macroscopic magnetization for $T<T_c$. However, this conclusion is apparently contradictory to observations. Consider the following example to explain this. If we want to magnetize a ferromagnetic material, say an iron needle, we either have to rub it with a permanent magnet in a unidirectional manner, or we need to put it in a strong magnetic field for a sufficiently long time (even when $T<T_c$). Only then the needle can be magnetized. The answer to this question was also given by Weiss. He argued that ferromagnetic materials contain many domains of microscopic size (known as Weiss domains). Each domain is big enough to contain large number of atoms with spins aligned in the same direction. In an unmagnetized  ferromagnetic material the directions of magnetization of the domains are randomly oriented with resultant zero magnetization. But when the sample is magnetized, domains rearrange to give a macroscopic magnetization.\\
 Pierre Weiss also designed large electromagnets and in 1918 established a Laboratory dedicated to magnetism in Strasbourg, France, where later in 1932 Louis Neel discovered anti ferromagnetism during his doctoral study. Neel was awarded with Nobel prize for this fundamental discovery in 1970.

\section{Pre-quantum mechanical era and the problems of the old quantum theory}
%—————————————————————————————————————————————————————————————————————————

The success and failure of the old quantum theory of Bohr and others are well known\cite{2}. And how the new quantum mechanics developed by Heisenberg, Born, Schroedinger, and Dirac replaced the patch-work of old quantum theory by a coherent picture of new quantum mechanics, in early 1920s, is also well known. In 1922, Stern-Gerlach experiment showed that magnetic moment of atoms can orient itself only in specific directions is space with respect to external magnetic field. This quantum mechanical phenomenon of spatial quantization was certainly missing in the Langevin treatment of paramagnetism. In the Langevin theory atomic moments can take any orientation in space. The required discretization of the spatial orientations was introduced, for the first time, by Pauli\footnote{Actually Pauli calculated electrical susceptibility. It turns out that same calculation goes through for magnetic susceptibility except one has to replace electric moment by magnetic moment\cite{3}.} who found that susceptibility expression with respect to the temperature variation is the same as that of Langevin but with different numerical coefficient $C$ in $\chi = C\frac{N \mu^2}{k_B T}$. He found the value 1.54 instead of 1/3 of the Langevin theory.  Pauli used integer quantum numbers but analysis of the band spectrum showed the need for half-integer values. Linus Pauling revised Pauli's calculation by using half-integer instead of integer values, and it resulted in another value of the coefficient $C$\cite{3}. The status of the field was far from satisfactory by 1925. There was another big problem. The calculations of susceptibility within the regime of old quantum theory appeared to violate the celebrated Bohr's correspondence principle, which states that in the asymptotic limit of high quantum numbers or high temperatures, the quantum expression should go over to the classical one ( as in black body radiation problem for $\frac{\hbar \om}{k_B T}<<1$). In the calculations of Pauli and Pauling there was no asymptotic connection with the Langevin theory. Then there was issues related to the weak and strong spatial quantization in the old quantum theory\cite{3}. Also the origin of the Weiss molecular field remained a complete mystery. {\it In conclusion, the old quantum theory of magnetism was a dismal failure. }

\section{Quantum mechanical and post-quantum mechanical era, and the development of the quantum theory of magnetism}
%—————————————————————————————————————————————————————————————————————————————————————————————————

The modern quantum mechanics was in place by 1926. The equivalence of the matrix formulation of Heisenberg (1925) and wave-mechanical formulation of Schroedinger (1926) by shown by Schoredinger in 1926. In the same period van Vleck attacked the problem of magnetism with "new" quantum mechanics.

\section{Enter van Vleck}
%%%%%%%%%%%%%%%%%%%%%%%

One of the pioneer of the quantum theory of magnetism is van Vleck who  showed how new quantum mechanics could rectify the problems of the old quantum mechanics, and restored the factor of 1/3 of the Langevin's semi-classical theory. In doing so he took space quantization of magnetic moment into account (instead of the integral in the partition function, proper summation was performed). In one of the pioneer investigation, van Vleck undertook a detailed quantum mechanical study of the magnetic behavior of gas nitric oxide $(NO)$. He showed quantitative deviations from semi-classical Langevin theory in this case, and his results agreed very well with experiments\cite{4}. The quantum mechanical method was applied to other gases, and he could quantitatively account for different susceptibility behavior of gases  like $O_2, NO_2$, and $NO$.\footnote{For a detailed account refer to his beautifully written book\cite{3}.} The differences in magnetic behavior arise from the comparison of energy level spacings ($\hbar \om_{if}$) with the thermal energy $k_B T$. He showed that the quantum mechanical expression for susceptibility reduces to the semiclassical Langevin result when all energy level spacings are much less than the thermal energy ($|\hbar\om_{ij}|<<k_B T$).  In the opposite regime (when for all $|\hbar\om_{ij}|>>k_B T$ )  $\chi$ showed temperature independent behavior. In the intermediate regime ($|\hbar\om_{ij}|\sim k_B T$) susceptibility showed a complex behavior (the case of nitric oxide). Thus van Vleck re-derived the Langevin theory by properly taking into account the space quantization.

Another major contribution of van Vleck is related to magnetism in solid-state. When  a free atom (suppose a free iron atom) becomes a part in a large crystalline lattice (like iron oxide), its energy levels change. The change in the electronic structure of an atom is due to two factors (1) outer electrons participate in the chemical bond formation, thus their energy levels change, and (2) in a crystalline lattice, the remaining unpaired electrons in the outer shells of an atom are not in a free environment, rather they are acted upon by an electrostatic field due to electrons on neighboring atoms. This field is called the crystalline field.     

Van Vleck and his collaborators introduced crystalline field theory (also known as the ligand field theory in chemical physics departments) to understand magnetic behavior in solid-state. With crystalline field ideas they could understand different magnetic behaviors of rare earth salts and iron group salts. It turns out that in rare earth salts $4f$ electrons are sequestered in the interior of the atom, and do not experience the crystalline field very strongly. The energy level splitting due to crystalline electric field is small as compared to thermal energy ($k_B T$), and it remains small even at room temperatures. Due to this the magnetic moment of the atoms behave as if the atom is free and shows the Langevin-Curie behavior $\chi\sim\frac{1}{T}$\cite{4,5,6}.

\begin{figure}[!h]
\begin{center}
\includegraphics[height=3cm]{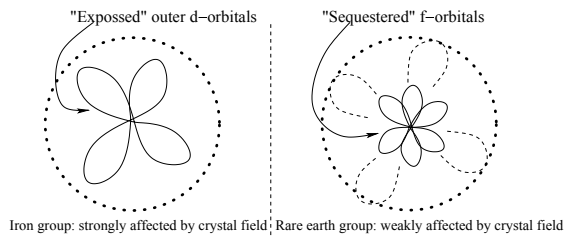}
\caption{A cartoon showing why crystal field effects differently an iron group ion and a rare earth ion.}
\end{center}
\end{figure}

In contrast to this case, in the iron group salts crystalline field is so strong that it quenches a large part of the orbital magnetic moment, even at room temperatures, leaving mainly the spin part to contribute to magnetism of salts of iron.

Magnetism of iron group {\it metals} is a different story (as compared to salts). In this case it turns out that charge carriers are also responsible for magnetism. The magnetism due to {\it itinerant electrons} was developed by Bloch, Slater, and Stoner \cite{nav}. The other extreme of localized electrons was investigated by Heisenberg. Van Vleck advanced ideas that can be dubbed as "middle of the way" approach (refer to \cite{nav}). For his pioneering contributions van Vleck was awarded with the Nobel  prize in physics in 1977 along with Phil Anderson and Nevill Mott. His articles are beautifully written and extremely readable and should form an essential element in a course (graduate or undergraduate) on magnetism. One can say that van Vleck is the father of the modern theory of magnetism, and his name will be forever remembered.

\section{Enter Dorfman}

When quantum mechanical study of magnetism of real gases was started by van Vleck in mid 1920s, the quantum mechanical study of magnetism in metals also started in the other continent transatlantic. 

The discovery of the paramagnetic properties of conduction electrons in metals is generally attached to Wolfgang Pauli. Pauli's paper came in 1927. Even before that, in 1923, Russian physicist Yakov Grigor'evich Dorfman put forward the idea that conduction electrons in metals posses paramagnetic properties\cite{7}. His proposal was based on a subtle observation: when one compares susceptibility of a {\it diamagnetic} metal with its ion, the susceptibility of the ion is always greater than its corresponding metal. It implies that there is some positive susceptibility in the case of the diamagnetic metal that partly cancels out the larger negative diamagnetic susceptibility. And this cancellation is prohibited in the case of metal's ion (due to ionic bonding). It was Dorfman's intuition that some positive susceptibility is to be attributed to conduction electrons in the metal i.e., some paramagnetic susceptibility has to be there. \footnote{It is important to note that the notion of the electron spin came in 1925 with a proposal by Uhlenbeck and Goudsmit and paramagnetism due to electron spin was discovered in 1927 by Pauli as mentioned before. But Dorfman's proposal came in 1923!} Dorfman's conclusion is based on his careful examination of the experimental data. After the discovery of the electron spin, Pauli gave the theory of paramagnetism in metals due to free electron spin. However, Dorfman was the first to point out paramagnetism in metals\cite{7}.

One of the other important contributions of Dorfman is his experimental determination of the nature of Weiss molecular field responsible for ferromagnetism in the Weiss theory. It was believed that the Weiss field is of magnetic origin due to which spins align to give a net spontaneous magnetization. To determine whether the Weiss field is of magnetic origin or of non-magnetic origin, Dorfman passed beta-rays (a free electron beam) in two samples of nickel foils, one magnetized and the other unmagnetized. From deflection measurements he determined that Weiss field is of non-magnetic origin\cite{8}.

In conclusion, Dorfman was an early contributor to the quantum theory of magnetism. But he is not  as well known as he should have been. 

\section{Enter Pauli}

Pauli's contribution to magnetism is well known. He formulated paramagnetic behavior of conduction electrons in metals in 1927 and showed that paramagnetic susceptibility is temperature independent (in the leading order). The derivation is discussed in almost all books devoted to magnetism and solid state physics\cite{9}. Pauli's derivation of the paramagnetic susceptibility can be described as one of the early application of Fermi-Dirac statistics of electrons in metals. In the standard derivation\cite{9} one calculates the thermodynamical potential $\Omega (H)$ of free electron gas in a magnetic field $H$.  Magnetization is obtained by the standard algorithm of statistical mechanics: $M = -\frac{\pr \Omega}{\pr H}$, and susceptibility $\chi = \frac{\pr M}{\pr H}$. For illustration purpose there is a simpler argument\cite{10} which goes like this. For metals at ordinary temperatures one has $k_B T<<E_F$ where $T$ is the temperature and $E_F$ is the Fermi energy. Thus electrons only in a tiny {\it diffusion zone} around the Fermi surface participate in thermodynamical, electrical, and magnetic properties (other electrons are paired thus dead). If $N$ is the total number of electrons, then fraction of electrons in the diffusion zone is $N\frac{T}{T_F}$ where $T_F$ is the Fermi temperature ($k_B T_F = E_F$).  Each electron in the diffusion zone has magnetic susceptibility roughly given by $\chi \sim \frac{\mu^2}{k_B T}$ where $\mu$ is its magnetic moment. Thus total magnetic susceptibility of metal is given by:  $N\frac{T}{T_F} \times \frac{\mu^2}{k_B T} = N\frac{\mu^2}{k_B T_F}$ which is independent of temperature as the more accurate calculation shows.

\section{Enter Heisenberg}

As mentioned before Dorfman in 1927 pointed out that the Weiss molecular field required in the theory of ferromagnetism is of non-magnetic origin. The puzzle of the Weiss molecular field was resolved by Heisenberg in 1928. The central idea is that it is the {\it quantum mechanical exchange interaction} which is responsible for the ferromagnetic alignment of spins. Quantum mechanical exchange interaction has no classical analogue, and it results due to the overlapping of orbital wave functions of two nearby atoms. Symmetry of the hybrid orbital is dictated by the nature of the spin alignment which obeys the Pauli exclusion principle. Thus there is an apparent spin-spin coupling due to orbital symmetry and under specific circumstances the ferromagnetic spin alignment significantly lowers the bonding energy thereby leading to a stable configuration.\footnote{It is very important to note that energy associated with spin-spin coupling of two electrons via exchange is very  large as compared to the magnetic dipole-dipole interaction energy which is given by \[V_{ij} = \frac{u_i.u_j}{r_{ij}^3} - 3 \frac{(u_i.r_{ij})(u_j.r_{ij})}{r_{ij}^5}.\] This very small magnetic energy cannot lead to ferromagnetic alignment. In other systems, like ferro-electrics it is an important energy.}\\

The Heisenberg model based on exchange interactions is related to the resonance-energy-lowering model for chemical bonding by Heitler and London\cite{11}.  In the Heitler-London theory of the chemical bond in hydrogen molecule, it is the exchange of electrons on two hydrogen atoms that leads to the resonant lowering of the energy of the molecule. Electrons stay in an antiparallel spin configuration  thereby enhancing the overlap of orbital wave functions in the intermediate region of two hydrogen atoms. This leads to bond formation.  This idea of resonant lowering of energy via exchange of electrons is greatly used by Linus Pauling in his general theory of the chemical bond\cite{11}. The Heisenberg model is built on similar ideas and goes like this\cite{5,12}. Let $S_i$ be the total spin at an atomic site $i$. If exchange interaction between nearest neighbors is the only one important, then the interaction energy (under certain approximations%\footnote{Here $S_i$ is the total spin at an atomic site $"i"$, i.e., it includes a vector sum over all the spins of unpaired electrons. In our notation $i$ and $j$ label two nearest sites.   Let $m$ and $n$ denote orbital numbers on a given site $i$ or $j$ (in cases where there are many unpaired spins in different orbitals).  Exchange interaction energy between an electron in $m$th orbital at site $i$ and an electron in $n$th orbital at site $j$ is given by \[V_{i,m;j,n} = -2 J_{i,m;j,n} S_{i,m} . S_{j,n}.\]  Total interaction is obtained by summing over all $m$ and $n$ \[V_{i,j} = -2 \sum_{m,n} J_{i,m;j,n} S_{i,m} . S_{j,n}.\]The main assumption is that the exchange integral between  $m$th orbital at site $i$ and $n$th orbital at site $j$ is {\it assumed to be independent of $m$ and $n$. It is like assuming the same exchange integral between two $s$-orbitals or two $d$-orbitals or between  $s$ and $d$ orbitals on two different sites $i$ and $j$. That is}\[J_{i,m;j,n}\simeq J_{i,j} \simeq J.\]  Validity of this assumption depends crucially on the nature of the system under consideration.  Of course, overlap of two S-orbitals is different  from that of two d-orbitals. But let us accept this assumption. Under this assumption $V_{i,j} = -2 J_{ij}S_i.S_j$ where $S_i = \sum_n S_{i,n}$ etc. Hence one obtains the Heisenberg model as given in the main text.})
is given by

\[V_{ij} = -2 J_{ij} S_i.S_j.\]
$J_{ij}$ is called the exchange integral\footnote{\[J_{ij} = \int d\tau_1\int d\tau_2 \phi_i(1)\phi_j(2) H_c \phi_j(1)\phi_i(2).\]}.  For ferromagnetism the sign of $J_{ij}$ has to be positive, and for anti-ferromagnetism it has to be negative. The question on what parameters the sign of $J$ depends is complicated and vexed one \cite{nav}. The above exchange interaction is now known as the Heisenberg exchange interaction or the direct exchange interaction. There is a variety of exchange interactions (both in metals and insulators) that are discussed in \cite{nav}. \\ 

To compare predictions of the model with experiment, one needs its solution. The very first solution provided by Heisenberg himself is based on some very restrictive assumptions.  So tight agreement with experiments may not be expected, and it leads to some qualitative results. Heisenberg used complicated group theoretical methods and a Gaussian approximation of the distribution of energy levels to find an approximate solution.\footnote{An alternative and comparatively simpler method was provided by Dirac using the vector model with similar conclusions\cite{1,2}.} From his solution Heisenberg observed that ferromagnetism is possible only if the number of nearest neighbors are greater than or equal to eight ($z=8$). This conclusion is certainly violated as many alloys show ferromagnetism with $z=6$. The second result which is much more important is that of magnitude of $\lambda$ it turns out that $\lambda$ of the Weiss molecular field takes the form

\[\lambda = z\frac{J}{2 N \mu_B^2}.\]
{\it The large value of $\lambda$ required for ferromagnetism is not a problem anymore, as the exchange integral $J$ can be large, thus resolving the problem of Weiss theory. This is the biggest success of the Heisenberg model. }\\

In conclusion, Heisenberg's model resolved the puzzle of the Weiss molecular field using the concept of exchange interaction. This concept turns out to be the key to the modern understanding of magnetism in more complex systems. Heisenberg's solution was based on many drastic assumptions which were later improved upon. Literature on the Heisenberg model and its various approximate solutions is very vast. Some references are collected here\cite{5,6,12,13}.

%\section{Attempts to solve the Heisenberg hamiltonian}

\section{Enter Landau}

Metals which are not ferromagnetic show two weak forms of magnetism, namely, paramagnetism and diamagnetism. Paramagnetism we have discussed, diamagnetism due to free conduction electrons is a subtle phenomenon and was a surprise to the scientific community\cite{1} when Lev landau discovered it in 1930.  To appreciate it consider the following example. Consider the classical model of an atom in which a negatively charged electron circulates around a positive nucleus. A magnetic moment  will be associated with the circulating electron (current multiplied by area).  Let  a uniform magnetic field be applied perpendicular to the electrons orbit. Let the magnitude  of the magnetic field be increased from zero to some finite value. Then, it is an easy exercise  in electrodynamics to show that an electromotive force will act on the electron in such a manner that will try to oppose the increase in the external magnetic field (i.e., Lenz's law). The induced opposing current leads to an induced magnetic moment in the opposite direction to that of the external magnetic field, and the system shows a diamagnetic behavior (induced magnetization in the opposite direction to the applied magnetic field).\\

However, when a collection of  such classical model-atoms is considered the diamagnetic effect vanishes. The net peripheral current from internal current loops just cancels with the opposite current from the skipping orbits (refer, for example, to \cite{1}). This observation also agrees with the Bohr-van Leeuwen theorem of no magnetism in a classical setting. Thus in a classical setting it is not possible to explain the diamagnetic effect.\\

However, in 1930, Landau surprised the scientific community by showing that free electrons show diamagnetism which arises from a quantum mechanical energy spectrum of electrons in a magnetic field. As described in many text books\cite{9} the solution of the Schroedinger equation for a free electron in a magnetic field is similar to that of the solution of the harmonic oscillator problem. There exits  equally  spaced energy levels - known as Landau levels. Each Landau level has macroscopic degeneracy. Statistical mechanical calculation using these Landau levels shows that there is non-zero diamagnetic susceptibility associated with free electrons which is also temperature independent as Pauli paramagnetism is.  And as is well known Landau level physics plays a crucial role in de Haas - van Alphen effect and related oscillatory phenomena, and in quantum Hall effects.\\ Our historical survey ends here. We will now discuss a current topics in magnetism that is the topic of spin liquids. There are other important developments in the field of magnetism, some of them are discussed in \cite{nav}.

\section{Spin Liquids}
We have studied the traditional magnetic orders (such as Ferromagnetism, paramagnetism etc) in which one can visualize and define magnetic order (like in ferromagnetism, all the spins are aligned in a particular direction and in antiferromagnetism one has two sub-lattices with oppositely directed spins). These magnetic orders can be incorporated into the general scheme of Landau's order parameter theory of second order phase transitions\cite{chaikin}. 
In Landau's theory, order parameter is zero at $T>T_{c}$ (higher symmetry phase) and is non-zero at $T \leq T_{c}$. 
(lesser symmetry phase). This is the standard paradigm. However, there are magnetic transition which do not fall under this general scheme. In fact, in these transitions an order parameter cannot be defined in the Landau sense. These systems with elusive magnetic order are called Quantum Spin Liquids(QSL). We will consider two examples where spin liquids are thought to exist. One is the enigmatic case of underdoped high transition temperature cuprate superconductors. For example consider the case of the parent compound $La_{2}Cu O_{4}$. This is a Mott insulator [Mott insulator are different from band insulator. They have partially filled bands but are insulators due to strong Coulomb repulsion]. It turns out that $Cu$ is in the oxidation state of $Cu^{++}$ i.e, it has one un-paired electron in the hybrid $Cu3d_{x^2-y^2}$-$O2P\sigma$ orbitals. This forms a narrow band, and two electrons on the same $Cu$ atom leads to large Coulomb repulsion (the Hubbard $U$)  and it is not energetically favourable to have double occupancy at a $Cu$ site. Thus, the system forms a $2D$ spin-$\frac{1}{2}$ lattice with AFM arrangement. Historically, it was point out by Anderson that in $2D$, strong quantum fluctuations will disrupt the long range $2D$ AFM arrangement and the state that will be realized is called RVB( Resonating Valence Bound state). In RVB, spin on nearest neighbours from singlet pairs and these pairs "resonate" between various configurations. Anderson's original proposal was along Pauling's theory of aromatic molecules in which a double bond resonates over single bonds, for example in Benzene. This quantum mechanical resonance lowers the energy of the system.
It turns out that Anderson's original proposal of RVB in $La_{2}Cu O_{4}$ was not correct, the state realised was in fact AFM(as has been verified by neutron scattering and other probes). But when one dopes the system ($ La_{1-x}Sr_{x}CuO_{4}$), this long range order is melted away and leads to some sort of dynamical order with short range AFM correlations. This state in underdoped cuprates is called the pseudogap state. Exact nature of the electronic order in this phase is not known as there is no order parameter in Landau sense. However, Anderson's original insight of resonating singlet pairs may be true in this phase put it is open to questions. It may be that pairs are making and breaking and are resonating over a vast  number of configurations. How to mathematically formulate it is a difficult current open problem in magnetism and in condensed matter physics.\\

Our second example of Quantum Spin Liquid (QSL) is Herbertsmithite with chemical formula $Zn Cu_{3}(OH)_{6}Cl_{2}$. The magnetically active element is again $Cu^{2+}$ with spin $S=\frac{1}{2}$. In this case also ordering is frustrated. If forms what is calles Kagome lattice, with edge sharing triangles \cite{takashi}. The well defined magnetic order is not stabilized in this case also. And the spins form single pairs and resonate between various quantum mechanical configurations ( fundamental feature of a spin liquid ). In contrast to the pseudogap state in cuprates, we have concrete evidence of a quantum spin liquid in Herbertsmithite. There are two experiments that points to the existence of QSL in this system. One is the neutron scattering and the other is the NMR Knight Shift experiments. Here we first discuss the neutron scattering experiment. If we have a nice magnetic order (like FM order) then neutron scattering leads to sharp Bragg peaks. And if  there are short range magnetic correlations then we have diffuse Bragg peaks. But if we have a diffused neutron scattering intensity over a considerable part of the Brillouin zone, even at very low temperatures one can then say that no magnetic order is selected by the system and it remains into a spin liquid state. This is what observed in Herbertsmithite. Neutron scattering intensity remains diffuse even at very low temperatures---clear signature of a QSL.\\ 

Another clear signature of the presence of QSL in this system comes from NMR relaxation rate. To understand that we need to first understand the nature of excitations in a QSL. As is well known excitations in a ferromagnetic are magnons. These are the quanta of deformation in magnetisation much like phonons which are quanta of lattice vibrations. It turns out that magnons are not gapped i.e. these can be excited with vanishingly small energy but the corresponding magnons will be of very long wavelength. In contrast, excitations in a QSL are gapped. These are called spinons.\\

For example in $1D$ AFM chain, if we flip one spin, then it generates two "unhappy" bonds and these excitations can propagate along the chain. The excitation of a spinon requires a finite amount of energy thus their excitation spectrum is gapped. When NMR relaxation rate of $Cu$ nuclear spin is measured it shows very weak relaxation at very low temperatures because at very low temperatures (such that $k_{B}T<\Delta$, where $\Delta$ is spin gap) spinon excitations are not there and alignment of nuclear spins is not degraded. Whereas at high temperatures there is a plenty of spinon excitations in QSL and nuclear relaxation becomes faster, as nuclear spin alignment is relaxed by spinon excitations. Thus we observe that QSLs is not an academic topic but they do exist in nature.\\

\section*{Acknowledgement}
Authors dedicate this article to the memory of Prof. N Kumar from whom they learned this topic. AMJ thanks DST for J.C. Bose national fellowship. NS would like to thank his friend Nagaraj (PRL Library) to provide required literature.

%	REFERENCE LIST
%----------------------------------------------------------------------------------------

%----------------------------------------------------------------------------------------

\end{document}